# On Bell-Like Inequalities for Testing Local Realism

Donald A. Graft
donald.graft@cantab.net

**ABSTRACT**

Objections to the use of inequalities to address foundational issues are considered and shown to be invalid. The Clauser-Horne (CH) inequality is considered and interpreted in this light. It is shown that, applied correctly, the CH inequality can validly test aspects of locality. This paper establishes a firm methodological ground for a following paper analyzing a recent CH experiment.

## 1. INTRODUCTION

John Bell famously claimed that the predictions of quantum mechanics for the results of EPRB experiments are inconsistent with the predictions of local realistic theories [1]. I show elsewhere [2] that this inconsistency follows not from the logic of quantum mechanics itself but rather from an incorrect *application* of quantum mechanics. I argue that the outcomes of EPRB experiments bear upon the question of whether the joint prediction [2] can be validly applied to cases of *separated* measurement, and not the broader question of the validity of quantum mechanics. Nevertheless, regardless of what we believe the experiments are testing, our understanding of nature would have to be adapted to the outcomes of the experiments. The needed adaptations (such as accepting the nonlocality of nature) would have significant foundational implications, and so it is imperative that the analysis and interpretation of our experiments stand on solid ground.

To interpret the results of EPRB experiments Bell developed several inequalities that are argued to be satisfied by local realism but violated by the accepted quantum prediction (here I am careful to recognize that if the argument of [2] is accepted, the correct quantum prediction would instead be based on marginal probabilities and not joint probability). Since the development of Bell's original inequalities, further inequalities have been developed. Each inequality has advantages and disadvantages. For example, the commonly used CHSH inequality [3] has the disadvantage that it is subject to the so-called 'detection loophole', i.e., violations of the CHSH inequality can be produced by detection losses alone. However, the Clauser-Horne (CH) inequality [4] cannot be violated by detection losses alone. As is well-known, and as shown by simulations in this paper, a minimum level of detection efficiency is still required for a decisive experiment. All of the EPRB experiments performed until very recently have not been definitive, because the detection efficiency has been too low. Recently, due to advances in technology, the detection efficiency has improved to the point where a definitive result can be achieved when the experiment is interpreted using the CH inequality. The CH inequality is immune to the detection loophole, that is, violations cannot be produced by detection losses, and this is its major advantage over other inequalities. We find ourselves now on the threshold of being able to obtain definitive answers to our foundational questions. It is an exciting time in the study of the foundations of quantum mechanics.

Despite these advances and the exciting future that they enable, there remains a minority section of the foundations community that strongly asserts that the use of inequalities is not justified, and that their use cannot shed any light on the foundational issues. These arguments cannot simply be ignored or summarily dismissed if we want our conclusions to stand on solid ground. This paper therefore addresses these arguments and shows them to be invalid. The CH inequality is then analyzed and its interpretation is discussed. It is argued that the CH inequality *can* validly test aspects of locality. A following paper [5] analyzes one of the recent experiments testing the CH inequality and shows, consistent with the argument in [2], that the experiment *confirms* local realism. This paper is intended to establish a firm methodological ground for the following paper.

## 2. OBJECTIONS TO THE USE OF INEQUALITIES

A main claim of this paper is that the CH inequality can validly test aspects of locality. This claim is challenged by variants of the general notion that the use of inequalities is methodologically wrong or suspect. To establish a firm ground on which to proceed, these objections must first be dispelled. In this section I consider the most important such objections and show that they do not succeed in blocking use of Bell-like inequalities, such as the CH inequality.

### 2.1 Are inequalities irrelevant to physics?

The idea that inequalities are just mathematical tautologies that are irrelevant to physics retains traction in some circles. It is surprising that this idea persists in the face of simple considerations that show it to be misguided. Nobody claims that a bare mathematical tautology can have empirical content and thereby tell us anything about physics. For example, we would waste our time searching for triangles that violate the triangle inequality (the sum of the lengths of any two sides of any triangle is greater than or equal to the length of the third side). But this is not what is being done when we apply an inequality in the context of the EPRB experiments. A valid and useful argument based on an inequality combines the tautologous inequality with a hypothesis.

For example, as shown later in this paper, the CH inequality combines a tautologous 6-term numerical inequality with the hypothesis that neither side has any information about the other side's measurement angles and outcomes. Testing just the bare inequality would of course be pointless, but testing the combined inequality-plus-hypothesis amounts to a test of the hypothesis. How can it be irrelevant to talk about whether the sides appear to communicate? The inequality component of an argument can be tautologous without making the whole irrelevant. If that were not the case, we could never use Boolean logic (deduction) in our arguments, because Boolean logic is tautologous! Inequalities used correctly are simply tools or expedients that enable or facilitate the testing of a hypothesis.

Beyond these considerations, in a following paper [5] I describe simulations of the CH inequality tests that demonstrate empirically the relevance and usefulness of CH-inequality-based tests. The idea that inequalities are irrelevant to physics cannot be sustained. Bell's core argument is fine. It says simply that an inequality can be used to determine if the two sides are somehow sharing ostensibly private information. If a violation is validly observed then the question becomes whether the sharing is occurring via some classical leakage mechanism (or other loophole) or via quantum nonlocality. A committed local realist will want to say that after leakage and loopholes have been eliminated, no violation can occur, and we see in [5] that this is indeed the case for a recent experimental test based on the CH inequality.

### 2.2 Does Bell's derivation of his inequality use probability incorrectly?

In a well-known paper, E. T. Jaynes criticized Bell's derivation and application of his inequality [6]. The critique can be interpreted in several ways. A simple reading of it has Jaynes claiming that Bell made an error in the application of probability theory. For example, Gill writes [7]: "According to E. T. Jaynes, Bell's factorization was an improper use of the chain rule for conditional probability." Bell had given the following integral for a local realistic model:

$$P(AB|\alpha,\beta,\lambda) = \int P(A|\alpha,\lambda)P(B|\beta,\lambda)p(\lambda)d\lambda \qquad (1)$$

However, in the simple reading, Jaynes is taken as arguing that (1) is a basic error in the application of probability theory, and that the correct formulation must use conditional probability as follows:

$$P(AB|\alpha,\beta,\lambda) = \int P(A|B,\alpha,\beta,\lambda)P(B|\alpha,\beta,\lambda)p(\lambda)d\lambda \qquad (2)$$

The Bell inequality derivation cannot proceed from (2) and therefore no conclusions are possible about locality. Like the objection described in Section 2.1, however, this objection fails to acknowledge that an additional (physical) hypothesis is involved in moving from (2) to (1). We start with the *hypothesis* that each side has no knowledge of the other side's parameter and outcome. This is the hypothesis we wish to test. We can formalize this hypothesis as follows (the integration over $\lambda$ is taken):

$$P(A|B,\alpha,\beta) = P(A|\alpha)$$
$$P(B|A,\alpha,\beta) = P(B|\beta) \qquad (3)$$

$$P(A \mid \alpha, \beta) = P(A \mid \alpha)$$

$$P(B \mid \alpha, \beta) = P(B \mid \beta)$$

The following formula given by conditional probability is tautologically correct:

$$P(AB \mid \alpha, \beta) = P(A \mid \alpha, \beta) P(B \mid A, \alpha, \beta) \tag{4}$$

Now we incorporate our hypothesis by replacing terms in (4) from the equations (3) defining our hypothesis:

$$P(AB \mid \alpha, \beta) = P(A \mid \alpha) P(B \mid \beta) \tag{5}$$

We have obtained Bell's factorization and therefore we can derive the Bell inequality. The inequality will *test* our hypothesis.

A more nuanced reading may suggest that Jaynes was indeed aware that an additional hypothesis is involved but that the hypothesis is not a valid formulation of a locality condition, i.e., that Bell has confused logical inference with causal influence. The factorization itself is accepted and the argument shifts to a question about what factorization signifies. In Section 3 I address the question of what the factorization means for the CH inequality. I argue that the inequality tests sharing of information between the sides that is ostensibly private. Locality imposes constraints on the possibilities for sharing of information, so (indirectly) we may test for locality using the inequality.

We see then that derivation of the inequalities does not involve any improper use of probability. Readers interested in other views on Jaynes' critique, and especially more detailed treatment of Jaynes' point about logical inference versus causal influence, may refer to Gill [7] and Potvin [8].

### 2.3 Are inequality violations due solely to incompatibility of observables?

Papers by Fine [9] and Malley [10] have inspired a view vociferously held in some circles that violation of the Bell inequality shows only that incompatible (noncommutative) observables have been illegitimately combined and that therefore the inequality can tell us nothing about locality. See Khrennikov [11] for one typical exposition of this view. An alternative statement of this position often heard is that derivation of the Bell inequality requires the existence of a single sample space for all the experiments, or that a joint PDF for all the experiments must exist. Violations then are attributed simply to the fact that such a single sample space does not exist, due to incompatibility of observables.

The literature surrounding this viewpoint is technical and involves what many would view as metaphysical considerations, such as objective realism, how to treat counterfactuals, etc. While I believe suitable responses are available to throw this whole incompatibility objection into doubt, see for example Stairs and Bub [12], I will here avoid entering into that debate and instead steer a simpler path to overcome the incompatibility objection and show that the inequalities are legitimate and can tell us something about locality.

Consider two incompatible observables A and B. (Note that incompatibility is not a concept limited to quantum mechanics; classical systems too can have incompatible observables.) An experiment measuring A (or B) alone will produce meaningful and useful results, but an experiment attempting to measure A and B together will not produce fully meaningful or useful results. That is what it means for the two observables to be incompatible. Let us now assume that we perform separated (marginal) measurements of A and B and have in hand two nonnegative numbers representing our measurement results that we call *P* and *Q*. Consider now the simple numerical inequality for any two nonnegative numbers that anybody may prove in seconds:

$$m + n \geq m \tag{6}$$

The inequality is a mathematical tautology; it must be true for any two nonnegative numbers (although here *m* may be negative; if we consider also the inequality $m + n \geq n$, we then require both *m* and *n* to be nonnegative). Now we simply insert our two measurement results *P* and *Q*, both legitimately obtained, into the tautology (6) to produce the following inequality:

$$P + Q \geq P \tag{7}$$

This resulting inequality is perfectly valid and no illegitimate steps were taken to obtain it. We have done two things: (1) generated a set of numbers representing results of measurements that do not involve incompatibility (because we measure them separately and not jointly), and (2) inserted these numbers into a tautologous inequality that is necessarily true regardless of the numbers inserted into it. What we must recognize here is that combining results in an inequality is *not the same* as trying to sample from a joint distribution that does not exist (due to incompatibility of the observables).

We now generalize this analysis to the CH inequality. We grant that a single joint PDF cannot be formed for all of the experimental arrangements (however, see below), but that is not required for derivation of the CH inequality. As we will see, derivation of CH requires only the specification of results from four experiments, *each of which involves only compatible measurements*, insertion of these results into a numerical tautology, and finally addition of a hypothesis about locality. Each experiment samples only a single PDF and so all of the experiments are individually valid. Only one Kolmogorovian sample space is used in any given experiment. There can thus be no objection to the CH inequality based on the notion of incompatibility, and it is easily seen that violation of CH is not isomorphic with performance of incompatible experiments.

As the incompatibility objection is often stated in the form of a claim that validity of the Bell inequality requires the existence of a single joint PDF for all the experiments, I comment further on this idea. One might reasonably argue that there *is* a single joint PDF that is parameterized. A parameterized joint PDF involves a set of sample spaces, one per context. Consider the quantum joint prediction $1/2\cos^2\theta$. Here we have a single functional form that is parameterized by the parameter $\theta$. This expedient of appealing to a parameterized distribution should be available to classical systems just as it is to quantum mechanics. Indeed, I have previously demonstrated a simple classical system generating quantum statistics that relies on a physical disk partitioned according to a parameter $\theta$ [2]. Again we have a single functional form, specific instances of which are instantiated for each experiment. For my cited model, it is as if we hold an infinite set of disks and $\theta$ defines which disk is used for a given experiment. I offer this idea of parameterized PDFs not to avoid the incompatibility objection, as I have shown that incompatibility is not involved in the CH inequality, but rather to suggest that strident declarations based on the nonexistence of a single distribution start from a needlessly constrained point of view. That is a separate debate, however, whose outcome does not affect my main argument.

Finally, of course, it is possible for contextuality to lead to violation of some other inequalities, which embody hypotheses that are not satisfied by the contextual conditions. For example, I and others have published models violating Bell's original 3-term inequality through contextuality. But the CH inequality cannot be so violated. Every inequality involves a tautology and one or more additional hypotheses. Care must be exercised to ensure that the additional hypotheses correspond only to the aspects of physical locality that we seek to test, and not to conditions unrelated to physical locality.

### 2.4 Can't the inequalities be violated by simple stochastic fluctuations?

Consider a system at the maximum of the Bell metric. The outcome streams can be corrupted by noise. A single bit flip can violate the Bell inequality for a given run. In practice, many bits are affected and we see a kind of random walk around the maximum of the Bell metric. One half of all runs will violate Bell's inequality. How can we distinguish such stochastic violations from real violations?

We might hope that the violation predicted by quantum mechanics is much larger than that due to stochastic noise, but this applies only for the ideal case of perfect detection and maximally entangled states. Current experiments deliver only about 75% detection efficiency and such levels require extreme nonmaximal states. The reduced efficiency and nonmaximal states combine to produce a severely attenuated metric, and it becomes difficult to distinguish a real effect from stochastic noise using only the magnitude of the violation. This diminution of the metric is exacerbated further in experiments such as the Christensen *et al.* experiment, where there is a large number of trials (Pockels cell openings) where no detection events occur, and so normalizing by the large number of trials dramatically reduces the measured CH metric. It's clear that some kind of metric normalized in a different way is needed that is not subject to reduction by the effects described, and that is useful for comparing simulations and experiments run under different conditions.

One possible approach takes advantage of the fact that for stochastic variations the fraction of violating runs will always be close to 50%. A real effect would strongly bias this fraction, thereby becoming evident. For example, if we observed 80% of runs violating the CH inequality, it would be hard to attribute that to stochastic violations. To apply this test, we can start with a long uninterrupted experimental run and partition it into (say) 100 sequential runs. Our stochastic metric now is the fraction of violating runs. It's easy to violate CH on a single run stochastically. The challenge is to violate it significantly more than half the time (half the runs).

Bierhorst [13] argues against use of this 50% rule because he claims it can be spoofed by particular temporal mixtures of source states, i.e., local models with appropriate source state distributions can be demonstrated that significantly violate the 50% rule. He argues that different statistical analysis is needed to exclude this loophole. However, the mechanism of this loophole applies only to an inequality *derived* from the bare CH inequality (see section 3) using additional hypotheses. The mechanism *does not violate* the bare CH inequality. Simulations of the mechanism show that, while it can easily obtain 60% or better violation when testing the derived inequality, only a 50% violation can be achieved when testing the bare CH inequality. Indeed, the mechanism *cannot* violate the bare CH inequality because no source state distribution can cause information to be transferred from one side to the other. The Bierhorst mechanism simply violates the additional hypotheses of the derived inequality. The locality hypothesis in the bare CH inequality is quite sufficient to test locality without needing additional, spurious hypotheses or derived inequalities. The 50% rule, then, remains fully valid for testing the bare CH inequality. Indeed, in [5] I show using the 50% rule that one recent CH-based experiment has been misinterpreted/misanalyzed and that its results confirm local realism.

The foregoing is not intended to argue that alternative statistical analyses cannot be valuable. For example, Gill [7] has employed martingales to assess the statistical significance of observed inequality violations. These alternative methods have been well-studied and the literature explores their different statistical strengths. I argue that the analysis with the 50% rule reveals very important things about the experiments, in fact, enough to confirm local realism [5]. Some of the things it reveals are: a) the amount of violation (of the 50% rule) depends on the maximality of the source state, b) the amount of violation depends on the detection efficiency, and c) the amount of violation depends on the size of the overall data set and how the data set is partitioned to apply the 50% rule. These behaviors can be plotted and compared to the predictions of local realism and quantum nonlocality. I argue that a simple metric that tells us a lot and is applicable to the actual experiments is fully adequate and should be preferred, and that maximization of the statistical strength of the analysis is unnecessary.

A sometimes overlooked point is that an experiment may show significant violation percentage when analyzed by partitions while the overall data set shows no violation. This could occur, for example, if several of the partitions have high noise. In a case like this we cannot even get started with the standard statistical test based on the number of standard deviations of violation, because there is no overall violation, and we would prefer to look at the percentage of partitions that violate the inequality. We could throw out the noisy partitions but that starts us on the slippery slope of data post-selection, and that should be avoided at all costs.

We see below that only information sharing between the sides can violate the 50% rule, and we see in [5] that simulations clearly demonstrate this, and we conclude that the CH inequality can tell us important things when assessed using the 50% rule.

### 2.5 Summary on the objections to use of inequalities

The general methodological objections to the use of inequalities have been dispelled. I now proceed to the derivation of the CH inequality abiding by the considerations adduced above, and to consider the meaning of the additional assumption used in the derivation.

## 3. DERIVATION AND INTERPRETATION OF THE CLAUSER-HORNE (CH) INEQUALITY

### 3.1 Derivation of the CH inequality

We follow the original derivation of Clauser and Horne [4]. Clauser and Horne derive the following inequality by inserting experimental results for marginal probabilities into a tautologous six-term numerical inequality:

$$P(A|\alpha,\lambda)P(B|\beta,\lambda) - P(A|\alpha,\lambda)P(B|\beta',\lambda) + P(A|\alpha',\lambda)P(B|\beta,\lambda) + P(A|\alpha',\lambda)P(B|\beta',\lambda)$$
$$- P(A|\alpha',\lambda) - P(B|\beta,\lambda) \leq 0 \tag{8}$$

Note that this inequality is still fully tautologous. The additional assumption to be tested has not yet been applied. Now we make an assumption of *factorizability*, i.e., the assumption that for any *a* and *b*:

$$P(AB|a,b,\lambda) = P(A|a,\lambda)P(B|b,\lambda) \tag{9}$$

We defer for the moment interpreting the meaning of this assumption and just say that we will seek to justify it with some physical considerations related to locality. Acceptance of the factorizability assumption allows us to replace terms in (8) to produce the following inequality:

$$P(AB|\alpha,\beta,\lambda) - P(AB|\alpha,\beta',\lambda) + P(AB|\alpha',\beta,\lambda) + P(AB|\alpha',\beta',\lambda) - P(A|\alpha',\lambda) - P(B|\beta,\lambda) \leq 0 \qquad (10)$$

Multiplication by ρ(λ) and integration over λ gives:

$$P(AB|\alpha,\beta) - P(AB|\alpha,\beta') + P(AB|\alpha',\beta) + P(AB|\alpha',\beta') - P(A|\alpha') - P(B|\beta) \leq 0 \qquad (11)$$

This is the CH inequality that we apply to the joint and marginal probabilities available from the four experiments involving the different combinations of measurement angles. Note that there are three other analogous inequalities that can also be derived:

$$-P(AB|\alpha,\beta) + P(AB|\alpha,\beta') + P(AB|\alpha',\beta) + P(AB|\alpha',\beta') - P(A|\alpha') - P(B|\beta') \leq 0$$

$$P(AB|\alpha,\beta) + P(AB|\alpha,\beta') - P(AB|\alpha',\beta) + P(AB|\alpha',\beta') - P(A|\alpha) - P(B|\beta') \leq 0 \qquad (12)$$

$$P(AB|\alpha,\beta) + P(AB|\alpha,\beta') + P(AB|\alpha',\beta) - P(AB|\alpha',\beta') - P(A|\alpha) - P(B|\beta) \leq 0$$

When we consider simulations and the data from actual experiments, we assess all four of these inequalities for possible violation. To eliminate simple stochastic violations, we can use the 50% rule described earlier, i.e., that significantly more than half the runs must show violations.

It is apparent that the derivation of the inequalities has proceeded mindful of the considerations adduced in Section 2, and that no methodological errors have been committed. The elephant in the room, of course, is the significance of the factorizability assumption, to which we now turn.

### 3.2 Significance of the factorizability assumption

The factorizability assumption (9) can be motivated and justified in various ways. Bell's own argument is that the assumption is justified by considerations of causality, light cones, and screening-off consistent with the Reichenbach common cause principle. While many accept and strenuously defend this analysis, others find it not fully persuasive due to its appeal to what may be viewed as metaphysical considerations. I find Bell's argument cogent but nevertheless believe that there is an alternative way to interpret factorizability that is not only more persuasive but also is more useful for interpreting the meaning of violations, as we will see. I discuss this *information causality* viewpoint [14][15][16][17] below, after considering the more conventional interpretation involving parameter and outcome independence.

Jarrett famously decomposed the factorizability condition into two subconditions: parameter independence and outcome independence (using Shimony's terminology in preference to Jarrett's). We previewed this decomposition in Section 2.2, where we used it to derive the factorizability condition. Parameter Independence (PI) is specified by:

$$P(A|\alpha,\beta) = P(A|\alpha)$$
$$P(B|\alpha,\beta) = P(B|\beta) \qquad (13)$$

Outcome Independence (OI) is specified by:

$$P(A|B,\alpha,\beta) = P(A|\alpha,\beta)$$
$$P(B|A,\alpha,\beta) = P(B|\alpha,\beta) \qquad (14)$$

The derivation of factorizability through the conjunction of PI and OI, however, leaves us feeling unsatisfied for several reasons. First, if a violation occurs, we don't know unambiguously whether it was due to a violation of PI, OI, or both, so the distinction may not be useful empirically. Second, the subconditions specify constraints only on the marginal probabilities, and so they may not capture everything that could be happening causally. For example, suppose we have a mechanism that is affected by a remote parameter C; the mechanism outputs 010101010101… when C is 0, and 001100110011… when C is 1. Here we have a case where the PI condition is satisfied, because the marginal probability for '1' is always 0.5, but where a clear causal mechanism allowing signaling is operating. Third, OI is equivalent to showing the conditional independence of *A* and *B* given the measurement angles and hidden variables. It is easy to establish the conditional independence for a deterministic system where we can be sure that the measurement angles and hidden variables fully specify the outcomes. However, for stochastic systems it is not so easy to establish the conditional independence. Finally, it's not at all clear how to relate this mess to actual locality conditions. For all these reasons, I eschew this decomposition and look toward the information causality approach for a clean intuitive interpretation of the factorizability condition.

The information causality approach concerns itself with the availability of information about parameters and outcomes, rather than with conditional marginal probabilities. Pawlowski *et al*. [14] derive from factorizability an inequality equivalent to the CHSH inequality but involving information transfer rather than conditional marginals. A similar information-based equivalent of the CH inequality can also be derived. Pawlowski *et al*. prove that violation of the inequality requires information transfer between the sides that necessarily includes information about *both* the outcomes and the measurement settings. Variations of my own disk model [2] as well as simulations confirm that transfer of both outcome and setting information are indeed required to violate the CH inequality. Outcome information may be implicitly transferred via the hidden variables without requiring communication between the sides, while the freedom to choose settings requires that some settings information is transferred between the sides.

The information causality interpretation is satisfying but not fully dispositive. We can conceive of four possibilities for the conditions for an experimental test: (a) no outcome or setting information is available to the other side, (b) outcome information only is available to the other side, (c) setting information only is available to the other side, and (d) outcome and setting information is available to the other side. Violation of the CH inequality tells us that possibility (d) is the case. Nonviolation leaves open possibilities (a), (b), and (c). Quantum mechanics (strictly, the misguided application of the quantum joint PDF to separated measurement situations) violates CH and thus (d) is implied. So we can reason as follows about the results of a CH test. If the test violates CH then there is confirmation of quantum information transfer (after leakage and loopholes have been eliminated). If the test does not violate CH then it disconfirms quantum information transfer and is consistent with full isolation (a), which is all a local realist could ask for, but still allows for two strange forms of information transfer (b) and (c).

The information causality perspective addresses only the sharing or isolation of information; it does not directly concern considerations of locality, i.e., constraints imposed by space and time. As was the case for the model of separation and localization this author previously propounded [2], space and time enter only through their constraints on the availability of information. For example, suppose side A performs a measurement and shouts to side B revealing the measurement angle and outcome. The physical medium for sound propagation here provides the mechanism for information transfer. As long as side B is within hearing range, side B can easily forge results (without any real measurements) yielding quantum correlation. If side A and side B move too far apart, side B can no longer hear the shouts and so cannot generate quantum results. Space has constrained the possibilities for sharing of information. A similar effect can occur in time.

A more germane example tying constraints on information transfer to considerations of locality might proceed as follows. We separate the sides of the experiment spatially by a very large amount, such that the limited speed of light cannot allow information to be transferred in the required timeframe. In such a physical scenario, detectors respond to the source events before information from the other side could possibly reach them (assuming settings are changed frequently enough). Such a scenario can test Einsteinian locality, even though the CH inequality can test *directly* only information availability. But even in a local experiment where both sides are collocated, we can enforce barriers to information flow. Violations of CH in such experiments would show that information has been transferred, despite the barriers, and if leakage and loopholes are eliminated, one would need to seriously consider the possibility of quantum information transfer or some other new physics.

### 3.3 Summary on the significance of the CH inequality

The CH inequality validly measures or detects the transfer of information about outcomes and settings from one side of the experiment to the other side. CH validly enforces that each side's detections (not just the marginal probabilities!) are not affected by information about the remote outcomes and settings. The relationship of a CH test to locality is physically contingent, as there are many ways to instantiate and control information. Locality is connected to separation, which is connected to the possibilities for information exchange, which is a valid objective criterion. If an experiment violates CH (and leakage and loopholes have been eliminated), then the experiment tends to confirm quantum information transfer (or quantum nonlocality in a suitable physical arrangement) or some other form of information transfer using unknown physics. If an experiment fails to violate CH (and detection efficiencies are high enough and noise low enough), then quantum information transfer is ruled out.

The CH inequality is known to exclude the effects of unfair sampling. This is most easily seen in a derivation of CH following Eberhard [18], in which missed detections are explicitly included. This feature of the CH inequality makes it preferable to the CHSH inequality. Dubious additional assumptions, such as the 'no-enhancement' assumption, are no longer required due to the greatly improved detection efficiencies recently achieved.

## 4. CONCLUSION

The methodological soundness of Bell-like inequalities was demonstrated. Common arguments against the use of inequalities were refuted. The 50% rule was shown to be adequate and valuable for analyzing the results of experiments testing the CH inequality. The factorizability assumption in the derivation of CH was interpreted from the perspective of information causality. We conclude that the CH inequality, when correctly applied, can validly test aspects of locality.